\documentclass[prd,preprint]{revtex4}
\usepackage{amssymb}

\begin{document}

\title{Gaugino Condensation in Heterotic Fivebrane Background}

\author{Noriaki Kitazawa}
\email{kitazawa@phys.metro-u.ac.jp}
\affiliation{
Department of Physics, Tokyo Metropolitan University,
Hachioji, Tokyo 192-0397, Japan}

\date{\today}

\begin{abstract}
The gaugino propagator
 is calculated by explicitly considering the propagation
 of a heterotic string between two different points in space-time
 using the non-trivial world-sheet conformal field theory
 for the fivebrane background.
We find that
 there are no propagations of gaugino
 which is in the spinor representation
 of the non-trivial four-dimensional space
 of the fivebrane background.
This result is consistent with the arguments
 on the fermion zero-modes of the fivebrane background
 in the low-energy heterotic supergravity theory.
Furthermore,
 assuming the continuous limit to the flat space-time background
 at the place far away from the fivebrane,
 we suggest an effective propagator
 which is effective only at the place far away from the fivebrane
 in the flat space-time limit.
From the effective propagator
 we evaluate a possible gaugino pair condensation.
The result is consistent with
 the suggested scenario of the gaugino condensation
 in the fivebrane background
 in the low-energy heterotic supergravity theory.
\end{abstract}

\preprint{}

\vspace*{3cm}

\maketitle

\section{Introduction}
\label{sec:intro}

The quantum effect of gravity might be important in particle physics.
For example, it was suggested that
 the supersymmetry can be spontaneously and dynamically broken
 by the gravitino pair condensation
 in non-trivial space-time backgrounds\cite{Witten,KMP}.
It is interesting that
 this mechanism seems rather model independent
 in comparison with other mechanisms in field theory without gravity.
Further investigation of this possibility
 requires a consistent theory of quantum gravity.
Although string theory is a strong candidate of it,
 the dynamics of the fermion states in string theories
 has not yet been extensively studied.
Especially,
 there are very few works about the fermion dynamics
 based on the framework of the string world-sheet theory.
In this paper we discuss the dynamics of the gaugino
 in the fivebrane background of heterotic string theory
 using the technique of the world-sheet conformal field theory.

The fivebrane background
 is a non-trivial background of space-time metric and fields
 in heterotic and type II string theories\cite{Rey1,CHS1,CHS2}.
The half of the original supersymmetry is broken,
 and the space-time Lorentz symmetry reduces to SO$(5,1)$
 from SO$(9,1) \supset$ SO$(5,1) \times$SO$(4)$
 by the fivebrane background.
An interesting fact is that
 some special fivebrane backgrounds can be described
 by world-sheet conformal field theories
 as the solutions of string theories\cite{Rey2,AFK}.
We calculate the gaugino propagator
 in the fivebrane background of heterotic string theory
 perturbatively with respect to the string coupling
 without any low-energy approximations
 by using the world-sheet conformal field theory.

If we have a fermion propagator,
 or more precisely a fermion two point function,
 in some non-trivial backgrounds
 of the space-time metric and gauge fields,
 it may be possible to obtain a value of the fermion pair condensate
 which can be an order parameter of some symmetry breaking.
For example,
 in Euclidean SU$(N)$ gauge theory
 with one massless vector-like fermion
 in the fundamental representation of SU$(N)$,
 the zero-instanton sector in the path integral
 of the fermion two-point function gives Euclidean propagator,
 and the one-instanton sector gives
 the value of the fermion pair condensate
 which triggers chiral symmetry breaking.
Therefore,
 it is interesting to investigate the fermion two point function
 in non-trivial backgrounds.

The number of fermion zero-modes
 in the non-trivial four-dimensional space of fivebrane backgrounds
 in low-energy supergravity theories is well-known\cite{CHS2}.
In the fivebrane background of heterotic string theory,
 there are two independent fermion zero-modes,
 and all three fermions, gaugino, gravitino and dilatino,
 have the zero-mode components\cite{Katagiri-Kitazawa}.
If we believe the path integral formalism of the supergravity theory,
 this number means the gaugino pair condensation
 and no propagations of gaugino which is
 in the SO$(4)$ (local Lorentz) spinor representation
 in the non-trivial four-dimensional space
 of the fivebrane background.

The paper is organized as follows.
In the next section
 we give a brief review of the fivebrane background
 in the heterotic string theory.
The world-sheet conformal field theory
 for the fivebrane background is introduced.
In Sec.\ref{sec:propagator} the gaugino propagator is calculated.
The {\it point boundary state}
 which couple with a single gaugino state is introduced.
The gaugino propagator is nothing but the transition amplitude
 between point boundary states of different places in space-time
 by the propagator operator in the world-sheet conformal field theory.
The method of the calculation is the same in Ref.\cite{Kitazawa}.
In Sec.\ref{sec:condensation}
 we discuss the limit of the fivebrane background
 to the flat space-time background
 at the place far away from the fivebrane.
Based on the discussion,
 we suggest an effective gaugino propagator
 which continuously reduces to the ten-dimensional free propagator
 in the flat space-time limit.
The effective propagator is effective
 only the place far away from the fivebrane.
By using this effective propagator
 the gaugino pair condensation at the place far away from the fivebrane
 is calculated.
In the last section we summarize our results.

\section{Fivebrane background in heterotic string theory}
\label{sec:fivebrane}

The fivebrane background (or the NS5-brane) is
 a Bogomol'nyi-Prasad-Sommerfield (BPS) configuration
 which preserves half of the supersymmetry of the theory.
In the heterotic supergravity theory, which is
 the low-energy effective theory of the heterotic string theory,
 the space-time metric and dilaton configuration
 of the fivebrane background (an NS5-brane at the origin)
 is explicitly given by
\begin{equation}
 g_{ab} = e^{2\Phi} \delta_{ab},
\qquad
 g_{{\bar \mu}{\bar \nu}} = \eta_{{\bar \mu}{\bar \nu}},
\label{metric}
\end{equation}
\begin{equation}
 e^{2\Phi} = e^{2\Phi_0} + {{n \alpha'} \over {r^2}},
\label{dilaton}
\end{equation}
where $a=1,2,3,4$ is the index
 of the non-trivial four-dimensional space,
 $r^2=x^a x^a$
 is the square of the distance from the fivebrane,
 and $n$ is some integer.
We use $\mu=0,1,\cdots,9$
 as the index of the whole ten-dimensional space-time,
 and use ${\bar \mu}=0,1,\cdots,5$
 as the index of the flat six-dimensional Minkowski space-time
 in the whole ten-dimensional space-time.
The geometry of the space-time is $M^6 \times R \times S^3$
 with varying radius of $S^3$ from $\sqrt{n\alpha'}$ to infinity
 along with the value of the coordinate $r$ from $0$ to $\infty$.
The background configuration of the B-field strength and gauge field is
\begin{eqnarray}
 H_{abc} = - \epsilon_{abcd} \partial^d \Phi,
\label{H-field}
\\
 A^I_a = 2 {\bar \eta}^I{}_{ab} \partial^b \Phi,
\label{gauge-field}
\end{eqnarray}
 where ${\bar \eta}^I{}_{ab}$ is the 't Hooft $\eta$ symbol
 and $I$ is the index of the adjoint representation of SU$(2)$
 which is embedded in the original gauge group
 of the heterotic string theory
 (SO$(32)$ in this paper).
This solution in the low-energy supergravity theory
 is considered to be exact in the heterotic string theory.
The world-sheet conformal field theory for these backgrounds
 can be explicitly constructed in case of $e^{2\Phi_0}=0$.
In this case
 the space-time geometry is $M^6 \times W^{(4)}_k$,
 where $W^{(4)}_k$ is the four-dimensional group manifold
 of SU$(2)_k \times$U$(1)$ with Ka\v c-Moody level $k$.
The geometry of $W^{(4)}_k$ is again $R \times S^3$,
 but the radius of $S^3$ is fixed with $\sqrt{n\alpha'} = \alpha'/Q$,
 where $Q \equiv \sqrt{\alpha'/(k+2)}$.

In case of the flat space-time
 the world-sheet theory of heterotic string theory
 consists of ten free boson fields $X^\mu(z,{\bar z})$,
 ten free fermion fields ${\tilde \psi}^\mu({\bar z})$
 and thirty-two free fermion fields $\lambda^A(z)$,
 where $A=1,2,\cdots,32$.
This system has
 $(N,{\tilde N})=(0,1)$ superconformal symmetry
 with the central charges of
 $c_m=10+36/2=26$ and ${\tilde c}_m=10+10/2=15$
 which are cancelled by the ghost contributions of
 $c_g=-26$ ($bc$-ghost contribution) and
 ${\tilde c}_g=-26+11=-15$ ($bc$ and $\beta\gamma$-ghost contributions).
The non-trivial background $M^6 \times W^{(4)}_k$
 can be described by replacing unconstrained fields
 $X^\mu$ of $\mu=6,7,8,9$
 by the fields constrained on the group manifold of
 SU$(2)_k \times$U$(1)$.
Namely,
 the part of the world-sheet theory
 corresponding to the space-time coordinates of $\mu=6,7,8,9$
 is replaced by the combination of the SU$(2)_k$
 Wess-Zumino-Witten (WZW) model
 and the linear-dilaton theory.
The new part has $(N,{\tilde N})=(4,4)$ superconformal symmetry,
 and has the same central charge of the original part:
 $c^{4D}_m={\tilde c}^{4D}_m=6$,
 where we treat the fields $\lambda^A$ of $A=29,30,31,32$
 as if they are superpartners of holomorphic boson fields.

The holomorphic sector of new SU$(2)_k \times$U$(1)$ part
 is described by three SU$(2)_k$ bosonic currents $J_i(z)$
 ($i = 1,2,3$), one free bosonic current $J_4(z)=\partial X^{\mu=6}$
 and four free fermionic fields $\Lambda_a \equiv \lambda^{28+a}$.
These currents and fields satisfy
 the following operator product expansion.
(We set $\alpha'=2$ from now on in this section, for simplicity.)
\begin{eqnarray}
 J_i(z) J_j(z') &\sim&
  - { k \over 2} {{\delta_{ij}} \over {(z-z')^2}}
  + \epsilon_{ijl} {{J_l} \over {z-z'}},
\\
 J_4(z) J_4(z') &\sim& - {1 \over {(z-z')^2}},
\\
 \Lambda_a(z) \Lambda_b(z') &\sim& - {{\delta_{ab}} \over {z-z'}}.
\end{eqnarray}
The $N=4$ superconformal symmetry transformation is generated
 by the following energy-momentum tensor $T^{W^{(4)}_k}(z)$,
 supercurrents $G^{W^{(4)}_k}_a(z)$ and SU$(2)_n$ currents $S_i(z)$.
\begin{eqnarray}
 T^{W^{(4)}_k} &=&
  - {1 \over 2}
  \left(
   {2 \over {k+2}} J_i^2 + J_4^2
   - \Lambda_a \partial \Lambda_a + Q \partial J_4
  \right),
\\
 G^{W^{(4)}_k}_i &=&
  \sqrt{{2 \over {k+2}}}
   \left(
    J_i \Lambda_4 - \epsilon_{ijl} J_j \Lambda_l
    + {1 \over 2} \epsilon_{ijl} \Lambda_4 \Lambda_j \Lambda_l
   \right)
  - J_4 \Lambda_i - Q \partial \Lambda_i,
\\
 G^{W^{(4)}_k}_4 &=&
  \sqrt{{2 \over {k+2}}}
   \left(
    J_i \Lambda_i
  + {1 \over {3!}} \epsilon_{ijl} \Lambda_i \Lambda_j \Lambda_l
   \right)
  + J_4 \Lambda_4 + Q \partial \Lambda_4,
\\
 S_i &=&
  {1 \over 2}
  \left(
   \Lambda_4 \Lambda_i
   + {1 \over 2} \epsilon_{ijl} \Lambda_j \Lambda_l
  \right).
\end{eqnarray}
The background charge $Q$
 determines the gradient of the linear-dilaton background
 $\Phi=QX^{\mu=6}$.
The world-sheet field $X^{\mu=6}(z,{\bar z})$
 is called the Feigin-Fuchs field,
 and the theory of this field is called the linear-dilaton theory.
The value $Q=\sqrt{\alpha'/(k+2)}=\sqrt{2/(k+2)}$ is required
 to have the correct central charge of $c^{4D}_m=6$,
 and the Ka\v c-Moody level $n$ of SU$(2)_n$ is fixed to unity
 due to the relation of $n=c^{4D}_m/6$.
The anti-holomorphic sector has exactly the same structure,
 except for that the fermion fields $\Lambda_a$ are replaced by
 ${\tilde \Psi}_a \equiv {\tilde \psi}^{5+a}$.

The world-sheet theory of the $M^6$ part consists of
 six free boson fields $X^{\bar \mu}(z,{\bar z})$,
 six free fermion fields ${\tilde \psi}^{\bar \mu}({\bar z})$
 and twenty-eight free fermion fields $\lambda^{\bar A}(z)$,
 where ${\bar A}=1,2,\cdots,28$.
In the holomorphic sector
 the energy-momentum tenor $T^{M^6}(z)$ is
\begin{equation}
 T^{M^6} = - {1 \over 2}
               \partial X^{\bar \mu} 
               \partial X_{\bar \mu}
             + {1 \over 2}
               \lambda^{\bar A} \partial \lambda_{\bar A}.
\end{equation}
In the anti-holomorphic sector
 the energy-momentum tenor and supercurrent are
\begin{eqnarray}
 {\tilde T}^{M^6}
        &=& - {1 \over 2}
               {\bar \partial} X^{\bar \mu} 
               {\bar \partial} X_{\bar \mu}
             + {1 \over 2}
               {\tilde \psi}^{\bar \mu}
                {\bar \partial}
               {\tilde  \psi}_{\bar \mu},
\\
 {\tilde G}^{M^6} &=&
  {\tilde \psi}^{\bar \mu} {\bar \partial} X_{\bar \mu},
\end{eqnarray}
 respectively.
The $M^6$ part has $(N,{\tilde N}) = (0,1)$ superconformal symmetry.
The whole world-sheet theory
 has $(N,{\tilde N}) = (0,1)$ superconformal symmetry,
 and we take
 $T=T^{M^6}+T^{W^{(4)}_k}+T^g$,
 ${\tilde T}={\tilde T}^{M^6}+{\tilde T}^{W^{(4)}_k}+{\tilde T}^g$
 and
 ${\tilde G}={\tilde G}^{M^6}+{\tilde G}_4^{W^{(4)}_k}+{\tilde G}^g$
 as the currents of that symmetry,
 where
\begin{eqnarray}
 T^g &=& \left( \partial b \right) c - 2 \partial \left( bc \right),
\\
 {\tilde T}^g &=&
  ( {\bar \partial} {\tilde b} ) {\tilde c}
  - 2 {\bar \partial} ( {\tilde b}{\tilde c} )
  + ( {\bar \partial} {\tilde \beta} ) {\tilde \gamma}
  - {3 \over 2} {\bar \partial} ( {\tilde \beta}{\tilde \gamma} ),
\\
 {\tilde G}^g &=&
  - {1 \over 2} ( {\bar \partial} {\tilde \beta} ) {\tilde c}
  + {3 \over 2} {\bar \partial} ( {\tilde \beta}{\tilde c} )
  - 2 {\tilde b}{\tilde \gamma}
\end{eqnarray}
 are ghost contributions.
The super-Virasoro generators $L_n$, ${\tilde L}_n$ ($n \in {\bf Z}$)
 and ${\tilde G}_r$ ($r \in {\bf Z}+1/2$ in Neveu-Schwarz sector
 and $r \in {\bf Z}$ in Ramond sector) are defined by
\begin{eqnarray}
 L_n &=& \oint {{dz} \over {2 \pi i z}} z^{n+2} \ T(z),
\\
 {\tilde L}_n &=&
  - \oint {{d{\bar z}} \over {2 \pi i {\bar z}}}
          {\bar z}^{n+2} \ {\tilde T}({\bar z}),
\\
 {\tilde G}_r &=&
  - \oint {{d{\bar z}} \over {2 \pi i {\bar z}}}
          {\bar z}^{r+3/2} \ {\tilde G}({\bar z}).
\end{eqnarray}
The explicit form of $L_0$, ${\tilde L}_0$ and ${\tilde G}_0$
 for holomorphic Neveu-Schwarz and anti-holomorphic Ramond sector
 (NS-${\tilde {\rm R}}$ sector) will be used in the next section.

The modular invariant partition function,
 or the one-loop vacuum amplitude,
 is explicitly given in Ref.\cite{AFK} for even $k$.
We consider $k$ as some even number from now on. 
One important fact is that
 in addition to the usual universal
 Gliozzi-Scherk-Olive (GSO) projection,
 the {\it additional GSO projection} is required
 for the modular invariance.
The half breaking of the space-time supersymmetry
 by the fivebrane background is realized
 by the additional GSO projection in the world-sheet theory.

The additional GSO projection
 of the fivebrane background in SO$(32)$ heterotic string theory
 also realizes the gauge symmetry breaking
 SO$(32) \rightarrow$ SO$(28) \times$SO$(4)$.
The gauge boson state is
\begin{equation}
 \lambda^A_{-1/2} \lambda^B_{-1/2} | 0 \rangle_{NS}
 \otimes
 {\tilde \psi}^\mu_{-1/2} | 0 \rangle_{\widetilde{{\rm NS}}},
\label{gauge}
\end{equation}
 where $\lambda^A_{-1/2}$ and ${\tilde \psi}^\mu_{-1/2}$
 are creation operators in NS and $\widetilde{{\rm NS}}$ sectors.
In case that
 the world-sheet boson part of the state
 has equal SU$(2)_k$ spins of
 the holomorphic and anti-holomorphic sectors,
 the additional GSO projection is nothing but the projection
 by the four-dimensional world-sheet fermion number.
There are two four-dimensional world-sheet fermion numbers
 for each holomorphic and anti-holomorphic sector, $F$ and ${\tilde F}$, 
 and $F+{\tilde F}$ should be even in the additional GSO projection.
For even ${\tilde F}$, $\mu=0,1,\cdots,5$ in Eq.(\ref{gauge}),
 then $F$ should be even, and
 $A$ and $B$ should take $A,B=1,2,\cdots,28$ or $A,B=29,30,31,32$
 in Eq.(\ref{gauge}),
 corresponding to the gauge bosons of
 SO$(28)$ and SO$(4)$ in six-dimensional space-time, respectively.
In case of $\mu=6,7,8,9$, namely $\tilde{F}$ is odd,
 $F$ should be odd, and $A=1,2,\cdots,28$ and $B=29,30,31,32$
 in Eq.(\ref{gauge}).
These states correspond to four six-dimensional matter scalar fields
 in $(28,4)$ representation of SO$(28) \times$SO$(4)$.

The gaugino state is
\begin{equation}
 \lambda^A_{-1/2} \lambda^B_{-1/2} | 0 \rangle_{NS}
 \otimes
 | \bf{s}_+ \rangle_{\widetilde{{\rm R}}},
\label{gaugino}
\end{equation}
 where $\bf{s}_+$ denotes the ten-dimensional spin state
 with even ten-dimensional chirality
 which is required by the universal GSO projection.
In case of the four-dimensional chirality of $\bf{s}_+$ is even
 (six-dimensional chirality is also even),
 namely $\tilde{F}$ is even,
 $A$ and $B$ should take $A,B=1,2,\cdots,28$ or $A,B=29,30,31,32$
 in Eq.(\ref{gaugino}) for even $F$.
These states are corresponding to
 two six-dimensional gauginos of SO$(28)$ and SO$(4)$, respectively.
In case of the four-dimensional chirality of $\bf{s}_+$ is odd
 (six-dimensional chirality is also odd),
 namely $\tilde{F}$ is odd,
 $A=1,2,\cdots,28$ and $B=29,30,31,32$
 in Eq.(\ref{gaugino}) for odd $F$.
These states are corresponding to
 two six-dimensional matter fermion fields
 in $(28,4)$ representation of SO$(28) \times$SO$(4)$.
All these six-dimensional fermion states of Eq.(\ref{gaugino})
 form six-dimensional $N=2$ supermultiplets
 with the boson states of Eq.(\ref{gauge}).
It is easy to check that other states,
 graviton, gravitino, dilaton and dilatino states,
 also form supermultiplets in six-dimensional $N=2$ supergravity theory
 \cite{Nishino-Sezgin}.

\section{Gaugino propagator in fivebrane background}
\label{sec:propagator}

The propagator in the heterotic string theory
 can be calculated as the transition amplitude
 between two appropriate boundary states
 by the propagator operator in the world-sheet conformal field theory
 \cite{CMNP,Fainberg-Marshakov,Marshakov}.
This propagator can include some off-shell information\cite{CMNP}.
In order to obtain the propagator
 in the field theory limit (not the low-energy limit),
 the {\it point boundary state},
 which describes the state of the closed string
 shrinking to a point in space-time,
 should be used.
We introduce the following point boundary state
 for the gaugino propagator in the fivebrane background.
\begin{equation}
 | B(y,{\bf s}) \rangle^{AB}
        = | B_X^{M^6} (y^{\bar \mu}) \rangle
  \otimes | B_{\rm{FF}} (y^6) \rangle
  \otimes | C \rangle
  \otimes | \psi_{\bf s} \rangle^{AB},
\label{boundary}
\end{equation}
 where each factor corresponds to the contribution of
 the bosons of $M^6$ part, the Feigin-Fucks field,
 the boson part of SU$(2)_k$ WZW model
 and all the fermions, respectively.
The explicit form of the first three factors
 is given in Ref.\cite{Kitazawa}.
Here we just notice that
 $y^{\bar \mu}$ specifies a point
 in six-dimensional Minkowski space-time,
 $y^6$ specifies a point in the seventh space
 in $Q \rightarrow \infty$ limit
 and $| C \rangle$ describes the state of the closed string
 shrinking to the north pole of $S^3$ of radius $\alpha'/Q$.
The explicit form of the contribution of fermion fields is given by
\begin{eqnarray}
 | \psi_{\bf s} \rangle^{\bar{A}\bar{B}} =
 \lambda^{\bar{A}}_{-1/2} \lambda^{\bar{B}}_{-1/2}
 | 0 \rangle_{\rm{NS}}
 \otimes | {\rm s} \rangle_{\tilde{\rm R}},
\\
 | \psi_{\bf s} \rangle^{\hat{A}\hat{B}} =
 \lambda^{\hat{A}}_{-1/2} \lambda^{\hat{B}}_{-1/2}
 | 0 \rangle_{\rm{NS}}
 \otimes | {\rm s} \rangle_{\tilde{\rm R}},
\end{eqnarray}
 where ${\bar A}=1,2,\cdots,28$ and ${\hat A}=29,30,31,32$,
 corresponding to SO$(28)$ and SO$(4)$ gauginos, respectively.
Both ten and six dimensional chirality of these states are even.

The propagator is obtained by explicitly calculate
\begin{equation}
 P^{(A'B')(AB)}(\psi_{{\rm s}'},\psi_{{\rm s}}, y'-y)
 = {}^{A'B'}\langle B(y',{\rm s}') | D | B(y,{\rm s}) \rangle^{AB}, 
\end{equation}
 where $A$ denotes ${\bar A}$ or ${\hat A}$ and
 $D$ is the propagator operator for NS-${\tilde{\rm R}}$ sector:
\begin{equation}
 D = {{\alpha'} \over {4\pi}} {1 \over 2}
     \int_{|z| \leq 1} d^2z {1 \over {|z|^2}}
     {\tilde G}_0 z^{L_0} {\bar z}^{{\tilde L}_0}.
\end{equation}
The calculation is straightforward (see Ref.\cite{Kitazawa})
 and the result is the following.
\begin{eqnarray}
 &P^{(A'B')(AB)}&(\psi_{{\rm s}'},\psi_{{\rm s}}, y'-y) =
 -i {\sqrt{\alpha'} \over 2}
 {1 \over \sqrt{2}}
  \left( \delta^{A'A} \delta^{B'B} - \delta^{A'B} \delta^{B'A} \right)
\nonumber\\&&
 \cdot
 \int {{d^7q} \over {(2\pi)^7}}
  e^{i q_{\tilde \rho} (y'-y)^{\tilde \rho}}
  {\bar \psi}_{{\bf s}'}
   \left(
    q_{\tilde \sigma} \Gamma^{\tilde \sigma}
    + {Q \over {\alpha'}} {1 \over {3!}}
       i \epsilon_{ijk} \Gamma^i \Gamma^j \Gamma^k
   \right)
  \psi_{\bf s}
\nonumber\\&&
  \cdot
  \sum_{N_X=0}^\infty d(N_X)
  \sum_{l=0}^k \sum_{m=-\infty}^\infty
  \left( 2(k+2)m+l+1 \right)
  \sqrt{{2 \over {k+2}}} \sin \left( \pi {{l+1} \over {k+2}} \right)
\nonumber\\&&
  \cdot
  {1 \over
   {
   q^{\tilde \lambda} q_{\tilde \lambda}
   + \left( {Q \over {\alpha'}} \right)^2 + M_{N_X}^2
   + {4 \over {\alpha'}}
     \left( (k+2) m^2 + (l+1) m \right)
   }},
\label{propagator1}
\end{eqnarray}
 where $\Gamma^\mu$ is the ten-dimensional Gamma matrix,
 ${\tilde \rho},{\tilde \sigma},{\tilde \lambda}=0,1,\cdots,6$
 and $i,j,k=1,2,3$ with $\Gamma^i \equiv \Gamma^{6+i}$.
The mass $M_{N_X}^2=4N_X/\alpha'$
 depends on the bosonic level $N_X$
 and $d(N_X)$ is the degeneracy of the open string state
 with the bosonic level $N_X$ and zero fermionic level.
Since we are interested in the gaugino propagation,
 we set $N_X=0$ in the following.
Furthermore,
 since the ten-dimensional space-time interpretation
 of this world-sheet theory is possible
 only in large $k$ limit, we take large $k$ limit.
The result in the momentum space is very simple
 (the over all normalization is changed).
\begin{equation}
 P^{(A'B')(AB)}(\psi_{{\rm s}'}, \psi_{{\rm s}} ,q) = 
 {1 \over \sqrt{2}}
  ( \delta^{A'A} \delta^{B'B} - \delta^{A'B} \delta^{B'A} )
 {\bar \psi}_{{\rm s}'}
  {{
    q^{\hat \mu} \Gamma_{\hat \mu} 
    + {Q \over {\alpha'}} {1 \over {3!}}
       i \epsilon_{ijk} \Gamma^i \Gamma^j \Gamma^k
   }
   \over
   {
    q^{\hat \mu} q_{\hat \mu} + (Q / \alpha')^2
   }}
 \psi_{{\rm s}}.
\label{propagator2}
\end{equation}
We have no momenta in $S^3$ space,
 since we did not introduce a specific point in that space.
We fixed the string at the north pole of $S^3$,
 but the absolute place of the north pole could not be fixed
 under the SU$(2)_k$ symmetry.
As a result, the propagation in $S^3$ is not incorporated
 \footnote{It is possible to incorporate the propagation in $S^3$
           by fixing the string at the south pole in the final state.}.
The mass $(Q / \alpha')^2$
 comes from the non-trivial torsion
 of the fivebrane background\cite{Kiritsis-Kounnas}.

Now we consider the additional GSO projection.
As discussed in the end of the previous section,
 the additional GSO projection is nothing but
 the six-dimensional chirality projection.
The additional GSO projection requires
 to take $\psi_{\rm s}$ and $\psi_{{\rm s}'}$ satisfying
\begin{equation}
 {{1+\gamma_7} \over 2} \psi_{\rm s} = \psi_{\rm s},
\qquad
 {\bar \psi}_{{\rm s}'}{{1-\gamma_7} \over 2}
  = {\bar \psi}_{{\rm s}'},
\end{equation}
 where $\gamma_7 \equiv \Gamma^0 \Gamma^1 \cdots \Gamma^5$.
Then the propagator becomes
\begin{equation}
 P^{(A'B')(AB)}(\psi_{{\rm s}'}, \psi_{{\rm s}} ,q) =
 {1 \over \sqrt{2}}
  ( \delta^{A'A} \delta^{B'B} - \delta^{A'B} \delta^{B'A} )
 {\bar \psi}_{{\rm s}'}
  {{
    q^{\bar \mu} \Gamma_{\bar \mu} 
   }
   \over
   {
    q^{\hat \mu} q_{\hat \mu} + (Q / \alpha')^2
   }}
 \psi_{{\rm s}}.
\label{propagator3}
\end{equation}
Note that there are no propagators of the gaugino
 in SO$(4)$ (local Lorentz) spinor representations
 with $\Gamma^\mu$ of $\mu=6,7,8,9$ in the numerator.
This is consistent with the arguments in low-energy effective theory.
There are two SO$(4)$ spinor zero-modes
 in non-trivial four-dimensional space in the fivebrane background
 in the low-energy supergravity theory\cite{Katagiri-Kitazawa},
 and there should be no SO$(4)$ spinor propagations,
 if we believe the path integral formalism of supergravity theories.
The instanton configuration
 in the non-trivial four dimensional space of the fivebrane background
 is not the one which indicates the instanton effect
 (the tunnelling effect between topologically non-trivial vacua),
 but the real configuration in four-dimensional Euclidean space.
Therefore,
 the path integral in non-trivial four-dimensional space
 includes only ``one instanton sector''
 without ``zero instanton sector''
 and results no gaugino propagator.

We should keep in mind that 
 the calculation in this section is
 in the lowest order of the string perturbation theory.
Since the value of the string coupling
 $g_s \propto e^\Phi$ becomes large at the place of the fivebrane
 (see Eq.(\ref{dilaton}) with $e^{2\Phi_0}=0$),
 the obtained propagator is effective only far away from the fivebrane
 (say $r>\alpha'/Q$).

\section{Possible gaugino condensation in fivebrane background}
\label{sec:condensation}

Since the gaugino condensation is an off-shell effect,
 there are no standard techniques to calculate it
 in string theory.
One of the possible strategy is
 to extract the information of the pair condensation
 from the propagator which includes some off-shell effect.
This strategy is rather familiar in the field theory.
For example,
 consider the quark propagator in QCD in four-dimensional space-time
 (though confinement might result no propagators of quarks).
A typical form of the propagator is
\begin{equation}
 {\rm F.T.} \langle 0 | T q {\bar q} | 0 \rangle
 = i {{p^\mu \gamma_\mu + \Sigma(p^2)} \over {p^2 - \Sigma(p^2)^2}},
\end{equation}
 where F.T. means Fourier transformation
 and $\Sigma(p^2)$ is a function of $p^2$.
Even though the quark is massless at the tree-level,
 the non-perturbative effect
 gives a non-trivial function of $\Sigma(p^2)$
 and gives the constituent mass to the quark.
The quark pair condensation can be roughly evaluated as
\begin{equation}
 \langle 0 | {\bar q} q | 0 \rangle
 \simeq
 - {\rm tr} \int {{d^4p} \over {(2\pi)^4}}
   {\rm F.T.} \langle 0 | T q {\bar q} | 0 \rangle
 = - \int {{d^4p} \over {(2\pi)^4}}
   {{4i\Sigma(p^2)} \over {p^2 - \Sigma(p^2)^2}}.
\end{equation}
The asymptotic freedom of the gauge interaction
 gives a finite value of the condensation.
It is easy to see that
 the condensation like $\langle 0 | {\bar q} \gamma^\mu q | 0 \rangle$
 vanishes.

If we naively apply this strategy
 to the gaugino propagator of Eq.(\ref{propagator3}),
 we have no condensations.
This is consistent with the observation that
 there are no gaugino condensations in six-dimensions.
The propagator can not be used to calculate the gaugino condensation
 in non-trivial four-dimensional space,
 since the method of the world-sheet conformal field theory
 with the additional GSO projection
 is not appropriate to describe the dynamics
 in non-trivial four-dimensional space.
The string states
 which are selected by the additional GSO projection
 can be understood as the states in the supermultiplets
 in six-dimensional $N=2$ supergravity theory,
 but they can not be understood as the fields in four-dimensional space.
This fact causes some difficulty to understand
 the physics in the flat space-time limit:
 $k \rightarrow \infty$ or $Q \rightarrow 0$.
(Note that
 we have to go to infinitely far away from the place of the fivebrane
 to keep the perturbation on the string coupling effective.
 See Eq.(\ref{dilaton}) with $e^{2\Phi_0}=0$
 and the relation $\sqrt{n\alpha'} = \alpha'/Q$.
 This is the strong coupling limit in view of the whole theory.)

The flat space-time limit of the propagator of Eq.(\ref{propagator3})
 does not coincide with the propagator
 in the flat ten-dimensional space-time.
Because of the additional GSO projection,
 the half of the original states remain to be projected out.
The same happens in the partition function in Ref.\cite{AFK}.
The partition function in the flat space-time limit
 is the half of the one in the flat space-time.
These facts imply that
 the flat space-time limit do not give
 the heterotic string theory in flat space-time.
This is due to the topological nature of the fivebrane background.

On the other hand, in the low-energy effective supergravity theory,
 all the non-trivial field configurations
 reduce to the trivial ones in the flat space-time limit
 at the place far away from the fivebrane.
The local world-sheet dynamics
 also continuously reduces to the one in the flat space-time,
 since the world-sheet energy-momentum tensor and supercurrent
 continuously reduce to the one in the flat space-time.
Therefore,
 it is natural to consider that
 the physics continuously reduces to the one in the flat space-time
 {\it at least locally at the place far away from the fivebrane}.

If this observation is correct,
 the effective propagator,
 which is effective only far away from the fivebrane,
 can be identified to the propagator of Eq.(\ref{propagator2})
 without additional GSO projection.
We can recover the spectrum in flat space-time
 by leaving off the additional GSO projection.
The global property of the fivebrane background is neglected,
 and only the local world-sheet dynamics remains.
Recovering the momentum in $S^3$ which now has very large radius,
\begin{equation}
 P^{(A'B')(AB)}_{\rm{eff}}(q) = 
 {1 \over \sqrt{2}}
  ( \delta^{A'A} \delta^{B'B} - \delta^{A'B} \delta^{B'A} )
  {{
    q^\mu \Gamma_\mu
    + {Q \over {\alpha'}} {1 \over {3!}}
       i \epsilon_{ijk} \Gamma^i \Gamma^j \Gamma^k
   }
   \over
   {
    q^\mu q_\mu + (Q / \alpha')^2
   }}.
\label{propagator}
\end{equation} 
This propagator continuously reduces to the one in flat space-time
 in $Q \rightarrow 0$ limit.

Now we can calculate the gaugino pair condensation.
Following the strategy
 which has been explained in the beginning of this section,
 we have
\begin{equation}
 \langle {\bar \lambda}^{AB} \Gamma \lambda^{AB} \rangle
 \simeq - \int {{d^{10}q} \over {(2\pi)^{10}}}
          {\rm tr} \left( \Gamma P^{(AB)(AB)}_{\rm{eff}}(q) \right),
\end{equation}
 where $\Gamma$ is some products of $\Gamma$-matrices.
The calculation is straightforward,
 and we have only one possible gaugino condensation:
\begin{equation}
 \langle {\bar \lambda}^{AB}
  \Gamma^7\Gamma^8\Gamma^9
 \lambda^{AB} \rangle
 = - 496 \times 32 \times {{{\rm Vol}(S^9)} \over {(2\pi)^{10}}}
   \cdot {Q \over {8 \alpha'^5}},
\label{condensation}
\end{equation}
 where the first factor
 comes from the contraction of the gauge group indices,
 the second factor comes from the trace of the $\Gamma$-matrix,
 and ${\rm Vol}(S^9)$ is the volume
 of the nine-dimensional unit sphere.
We have set the ultraviolet cut off $1/\sqrt{\alpha'}$
 which is the natural cut off scale in the low-energy effective theory.
This result is consistent with the scenario of the gaugino condensation
 which is suggested in Refs.\cite{Katagiri-Kitazawa,GSW,GCDL}.
The Lagrangian of the heterotic supergravity theory\cite{BdR}
 contains a term
\begin{equation}
 {\cal L}_{\rm hetero}
  \supset - {1 \over {6 \kappa^2}} e e^{-2\Phi}
  \left(
   H_{\mu\nu\rho}
   - {\kappa^2 \over {16}}
     {\bar \lambda}^{AB} \Gamma_{\mu\nu\rho} \lambda^{AB}
  \right)^2,
\label{parfect-square}
\end{equation}
 where $\kappa \propto \alpha'^2$
 determines the gravitational scale in ten dimensions.
In the fivebrane background
 $H$ has the non-trivial value of Eq.(\ref{H-field}).
Therefore,
 the composite operator
 ${\bar \lambda}^{AB} \Gamma_{\mu\nu\rho} \lambda^{AB}$
 could be expected to condense
 for which the term of Eq.(\ref{parfect-square}) vanishes
 for the vanishing cosmological constant.
In the coordinate system of the present string world-sheet theory,
 only non-zero value of $H$ is $H_{789}=Q/\alpha'$.
Therefore,
 the condensation of Eq.(\ref{condensation})
 is totally consistent with this scenario.

This gaugino condensation may act an important role
 in the half supersymmetry breaking by the fivebrane background.
In low-energy effective theory,
 we know the half supersymmetry breaking
 by the fact that the fivebrane background configuration is invariant 
 under only the half of the original supersymmetry transformations.
In the algebraic point of view,
 the partial supersymmetry breaking
 must be accompanied by the central charge
 in the supersymmetry algebra\cite{Witten-Olive,Chibisov-Shifman}.
It is well known that the central charge
 is induced by the topologically non-trivial configuration
 of the boson fields in space-time.
There may be another contribution to the central charge
 through the quantum effect like the Konishi anomaly
 which gives a gaugino bilinear term
 in the supersymmetry transformation of some composite operators.
There might be a possibility that
 the bilinear term of the gaugino
 (and also of the gravitino and dilatino)
 appears in the supersymmetry algebra in heterotic supergravity theory,
 and its condensation gives a contribution to the central charge
 for the half supersymmetry breaking by the fivebrane background
 (I would like to thank S.-J.~Rey for pointing out this possibility).

\section{Summary}
\label{sec:conclusions}

In string theory it is difficult to obtain the off-shell information,
 like the fermion pair condensation.
The fermion propagator
 which may contain some information about the fermion pair condensation,
 can be concretely calculated by several methods in string theory,
 though no methods for the non-perturbative calculation are available.
Since the fermion pair condensation
 is expected to be a non-perturbative effect,
 it seems impossible to treat it in string theories.
But we know the instanton calculus in gauge theories
 in which the values of the condensates of fermion composite operators
 can be perturbatively calculated by the path integral
 in instanton backgrounds.
The existence of the instanton configurations as classical solutions
 is itself has a non-perturbative meaning,
 and the non-perturbative effect is incorporated
 by the semi-classical approximation. 
It is expected that
 the similar situation is realized in the string propagator
 in non-trivial backgrounds.

Some non-trivial backgrounds in string theories
 are exactly described by using the world-sheet conformal field theory,
 and the propagators can be calculated perturbatively.
The fivebrane background in the heterotic string theory
 is one of such background configurations,
 and the gaugino pair condensation
 in the non-trivial four-dimensional subspace of the space-time
 (and also the gravitino and dilatino pair condensations)
 is expected by the analysis in the low-energy supergravity theory.
We calculated the gaugino propagator in the fivebrane background,
 and found no propagation of the gaugino in SO$(4)$ (local Lorentz)
 spinor representation.
Although this result is consistent with the arguments
 about the fermion zero-modes in the low-energy supergravity theory,
 we can not obtain the value of the gaugino condensate.
The rigorous formulation for the fivebrane background
 allows to describe the dynamics only in six-dimensional space-time.

Based on the observation that
 in the flat space-time limit, $k \rightarrow \infty$,
 the physics at the place far away from the fivebrane
 should continuously reduce to the one in the flat space-time,
 we suggested that
 the effective propagator at the place far away from the fivebrane
 is just the calculated propagator
 without the additional GSO projection.
By using this effective propagator,
 we evaluated a possible gaugino pair condensate.
The result is perfectly consistent
 with the suggested scenario of the gaugino condensation
 in low-energy effective supergravity theory.

The meaning and role of this condensation are unknown.
One possibility is that
 the condensation contributes the central charge
 in the supersymmetry algebra in ten-dimensions.
The central charge is required
 to realize the half supersymmetry breaking by the fivebrane.

\acknowledgments

I would like to thank S.~Saito for helpful comments.
I also would like to thank S.V.~Ketov for encouragements.
This work was supported in part by Yamada Science Foundation
(No. 2003-4022).

\end{document}